\documentclass[12pt]{article}

\def\beq{\begin{eqnarray}}
\def\eeq{\end{eqnarray}}
\def\m{M_*}

\def\mpl{M_{\rm Pl}}
\def\e{{\epsilon}}

\def\d{\delta^{(N)}(y)}

\def\g{{\cal G}}

\newcommand{\gsim}{\lower.7ex\hbox{$\;\stackrel{\textstyle>}{\sim}\;$}}
\newcommand{\lsim}{\lower.7ex\hbox{$\;\stackrel{\textstyle<}{\sim}\;$}}

\begin{document}

\begin{titlepage}

\begin{flushright}
{ NYU-TH-03/12/23; CERN-TH/2003-158

FTPI-MINN-03/19, UMN-TH-2208/03}
\end{flushright}
\vskip 0.9cm

\centerline{\Large \bf  Softly Massive Gravity}

\vskip 0.5cm
\centerline{\large G.~Gabadadze ${}^{a,b}$ and M. Shifman 
${}^{b,c}$}
\vskip 0.2cm
\centerline{\em ${}^a$  Center for Cosmology and Particle Physics}
\centerline{\em Department of Physics, New York University, New York, 
NY, 10003, USA$^*$}
\vskip 0.2cm
\centerline{\em ${}^b$ Theory Division, CERN, CH-1211 Geneva 23, Switzerland}
\vskip 0.2cm
\centerline{\em ${}^c$ William I. Fine Theoretical Physics Institute,
University
of Minnesota,}
\centerline{\em Minneapolis, MN 55455, USA$^\diamond $}

\vskip 0.9cm

\begin{abstract}

\vskip 0.2cm

Large-distance modification of gravity 
may be {\em the} mechanism for solving the  
cosmological constant problem. A simple model of the 
large-distance  modification ---  four-dimensional (4D) gravity with 
the hard mass term--- is problematic from the theoretical standpoint. 
Here we discuss a different model, {\it the brane-induced gravity}, 
that effectively introduces a soft graviton mass.
We study the issues of unitarity, analyticity and causality in this model 
in more than five dimensions. We show that a consistent prescription for the poles 
of the Green's function can be specified so that 4D unitarity is preserved. 
However, in  certain instances 4D analyticity cannot be maintained 
when theory becomes higher dimensional. As a result, one has to sacrifice 
4D causality at distances of the order of the present-day Hubble scale.
This is a welcome feature for solving the cosmological constant 
problem, as was recently argued in the literature.
We also show that, unlike the 4D massive gravity,  
the model  has no strong-coupling  problem at 
intermediate scales.

\end{abstract}

\vspace{3cm}

\noindent
\rule{2.4in}{0.25mm}\\

\noindent
$^*$ {\small Present address} \\
$^\diamond$ {\small Permanent address}

\end{titlepage}

\section{Large distance modification of gravity: \\
formulating the problem}
\label {ldmogftp}

\noindent
The reason underlying the observed acceleration of the universe
is puzzling. It could be a tiny amount of
vacuum energy. However, this possibility is hard to reconcile
with  known particle-physics models. Instead, it might well
be that a new physical scale exists in the gravitational sector
and the laws of gravity and cosmology are modified
at this scale. To be consistent with  data  and be able
to predict the accelerated expansion, the new
scale should be roughly equal to
$H_0^{-1}\sim 10^{28}\,\,{\rm cm}$ --- 
the present-day value of the Hubble length. 
In this regard, developing  models in which  gravity 
gets modified at cosmological distances, becomes a timely endeavor.
A generally covariant theory  of the large-distance modification of gravity 
is the DGP model \cite {DGP}. The gravity action of the  
model can be written as follows:
\beq
S\,=\,{\mpl^2\over 2} \,\int\,d^4x\,\sqrt{g}\,R(g)\,+
{\m^{2+N}\over 2}\,\int \,d^4x\,d^Ny\,\sqrt{{\bar g}}
\,{\cal R}_{4+N}({\bar g})\,,
\label{1}
\eeq
where
$R$ and ${\cal R}_{4+N}$ are the four-dimensional
and $(4+N)$-dimensional Ricci scalars, respectively, and
$\m$ stands for the gravitational scale
of the bulk theory. Extra dimensions are not compactified,
they asymptote at infinity to Minkowski space.
The higher-dimensional and
four-dimensional  metric tensors are related as
\beq
{\bar g}(x,y=0)\equiv g(x)\,.
\label{bargg}
\eeq
The first term on the right hand side of (\ref {1}) acts as a kinetic term for a 4D 
graviton while the second term acts as a gauge invariant mass term.
The observable matter is assumed to be localized on a 4D surface $y=0$.

The present work is devoted to the study
of the DGP scenario in the case
$N\ge 2$ (see Ref.~\cite {DG}, \cite {MP}). Such models
have a string theory realization \cite {Ignat}. More importantly,
these models are potential candidates for  solving~\cite {DGS,Zura} the
cosmological constant problem
(see also Refs. \cite{Cedric}--\cite {Roman}
for interesting cosmological and astrophysical studies).

The equation
of motion for the theory described by the action 
(\ref {1}) takes the form 
\beq
\d \, \mpl^2 \, G^{(4)}_{\mu\nu}\,\delta^\mu_A\,\delta^\nu_B
\,+\, \m^{2+N} \,G^{(D)}_{AB}\,
= \, - T_{\mu\nu}(x)\, \delta^\mu_A\,\delta^\nu_B \, \d \,.
\label{geq0}
\eeq
Our conventions are as follows:
\begin{eqnarray}
\eta_{AB} &=& {\rm diag}\, [+--...-]\,,\qquad A,B = 0,1,...,3+N\,,
\nonumber\\[1mm]
\mu,\nu &=& 0,1,2,3\,,\qquad a,b=4,5,...,3+N\,.
\end{eqnarray}
$G^{(4)}_{\mu\nu}$ and $ G^{(D)}_{AB}$ 
denote the four-dimensional  and
$D$-dimensional Einstein tensors, respectively. 
We choose (for simplicity)  a source  localized on 
the brane, $T_{\mu\nu}(x)\d $.

Gravitational dynamics encoded in Eq.~(\ref {geq0})
can be inferred  both  from the four-dimensional (4D)
as well as $(4+N)$-dimensional standpoints.
 From the 4D perspective,  gravity on the brane is mediated
by an infinite number of the Kaluza-Klein  (KK) modes that
have no mass gap. Under  conventional circumstances
(i.e.,  with no  brane kinetic term) this would
lead to higher-dimensional interactions. However,
the large 4D Einstein-Hilbert (EH) term
suppresses the wave functions of  heavier KK modes,
so that in effect they  do not participate  in the
gravitational interactions  on the brane at observable distances
\cite {DGKN}.  Only  light KK modes,
with masses $m_{KK}\lsim m_c$,
\beq
m_c \equiv \frac{\m^2}{\mpl}\,,
\label{mc}
\eeq
remain essential,
and they collectively act  as an effective 4D graviton with
a typical mass of the order of $m_c$ and a certain smaller width.

Assuming that $\m\sim 10^{-3}$ eV
or so, we obtain  $m_c \sim H_0\sim 10^{-42}$ GeV.
Therefore, the DGP model with $N\geq 2$ 
predicts~\cite {DHGS} a modification of gravity 
at short distances $\m^{-1}\sim 0.1$ mm {\em and} at large 
distances  $m_c^{-1}\sim
H_0^{-1}\sim 10^{28}\,\,{\rm cm}$, give or take an order of magnitude.
Since gravitational interactions,  nevertheless, are mediated
by an infinite number of states at  arbitrarily low
energy scale,  the effective theory  (\ref {1}) presents, from the 4D 
standpoint, a {\it non-local} theory~\cite {DGS}. Moreover,
as was suggested in \cite {ADDG}, nonlocalities postulated in 
{\em pure  4D} theory  can solve an ``old'' cosmological constant problem \cite {ADDG}, 
and give rise to  new mechanisms  for the present-day acceleration 
of the universe~\cite{ADDG,Carroll}. (It is interesting to note that the nonlocalities 
in a gravitational theory that are needed to solve the cosmological constant problem 
could appear from quantum gravity \cite {Veneziano} or matter 
loops \cite {Mottola} in a purely 4D context.)

On the other hand, from the $(4+N)$-dimensional perspective,
gravitational interactions are  mediated by a single
higher-dimensional graviton. This graviton has two kinetic terms
given in Eq.~(\ref {1}), and, therefore, can propagate differently
on and off the brane. Namely, at short distances, i.e. at
$r< m_c^{-1} \sim H_0^{-1}\sim 10^{28}\,\,{\rm cm}$,
the graviton emitted along the brane
essentially propagates along  the brane and mediates
4D interactions. However, at larger distances, the
extra-dimensional effects take over and gravity
becomes ($4+N$)-dimensional.

As was first argued in Ref. \cite {DG}, the results in 
$N\ge 2$ DGP models are sensitive to  ultraviolet (UV) physics, 
in contradistinction to  the $N=1$ model \cite {DGP}. 
In other words, one should either consistently smooth out the 
width of the brane \cite {DHGS}, or introduce a 
manifest UV cutoff in the theory \cite {Wagner,Kiritsis,DHGS}, 
or do both. With a finite thickness,  more localized 
operators  appear on the worldvolume of the brane, 
in addition to  the worldvolume Einstein-Hilbert term 
already present in  Eq.~(\ref {1}) \cite {DHGS}. For instance, one could  
think of a higher-dimensional Ricci scalar 
smoothly spread over  the worldvolume \cite {MP}. 
 
In general, terms that are square of the extrinsic curvature can 
also emerge. Some of these
terms can survive in the limit when the brane thickness 
tends to zero (i.e. in the low-energy approximation).
For instance, in the zero-thickness limit  of the brane
the following terms might  be important:
\beq
\d\, h^\mu_\mu \,\partial_\alpha^2 \, h^a_a\,,~~~
\d\, h^{\mu\nu}\,\partial_\mu \partial_\nu \, h^a_a\,,~~~\d \,h^\mu_\mu
\, \partial_a \partial_b \, h^{a b}\,,
\label{terms}
\eeq
where $h$ denotes small perturbations on flat space.
Although the main features of the model, such as  
interpolation between the 4D power-law behavior of a non-relativistic 
potential at short distances and the 
higher-dimensional behavior at large distances,
are not expected to be changed by adding these terms, 
nevertheless, the tensorial structure of a propagator 
could in general depend on these terms and 
self-consistency of the theory may 
require some of these terms to be present in the actions in a 
reparametrization invariant way.

In the low-energy approximation the exact form of these ``extra" terms 
and their coefficients  are ambiguous, because of their UV origin.
They will be fixed in a fundamental theory 
from which the DGP model can be derived \cite {Ignat,AGV}.
In the present paper, in the absence of such a fundamental theory, 
(but in the anticipation of its advent),
we would like to study a particular parametrization of these ``extra" 
terms,
for demonstrational purposes. According to our
expectations, physics in the 
self-consistent theory will have   properties very similar 
to those discussed below. We will show that these properties are 
rather attractive since they do avoid severe problems of 4D 
massive gravity. 
 
 \vspace{2mm}
 
Consider the action 
\beq
S\,=\,{\mpl^2\over 2} \,\int\,d^4x\,\sqrt{g}\,\left ( a \,R(g)\,+\,
b \, {\cal R}_{4+N} \right )\,+
{\m^{2+N}\over 2}\,\int \,d^4x\,d^Ny\,\sqrt{{\bar g}}
\,{\cal R}_{4+N}({\bar g})\,,
\label{11}
\eeq
where, in addition to the 4D EH term,   a 
D-dimensional EH term localized on the brane is present. Here $a$ and $b$ are 
some numerical coefficients. 
We will study the properties of the system described by (\ref{11})
for different values of $a$ and $b$. The action (\ref{11})
is fully consistent with the philosophy of Ref.~\cite{DGP}:
if there is a (1+3)-dimensional brane in D-dimensional space,
with some ``matter fields" confined to this brane,
quantum loops of the confined matter will induce
all possible structures consistent with the
geometry of the problem, i.e. (1+3)-dimensional wall
embedded  in D-dimensional space.

The equation of motion in the  model (\ref {11}) takes 
the form 
\beq
\d \, \mpl^2 \, \left ( a\,G^{(4)}_{\mu\nu}\,+\, b\,G^{(D)}_{\mu\nu} \right )
\delta^\mu_A\,\delta^\nu_B
\,+\, \m^{2+N} \,G^{(D)}_{AB}\,
= \, - T_{\mu\nu}(x)\, \delta^\mu_A\,\delta^\nu_B \, \d \,.
\label{geq}
\eeq
In deriving the above equation we first 
introduced a finite brane width $\Delta$, 
and then took the $\Delta \to 0$ limit in such a way 
that  no surface terms appear. In general, the results 
depend on the regularization procedure for the brane width. 
In the present work we adopt a simple prescription in which 
derivatives with respect to the transverse coordinates calculated 
on the brane vanish in the $\Delta \to 0$ limit (a unique prescription 
could only be specified by a fundamental theory.).
As previously,  $G^{(4)}$ and $ G^{(D)}$ 
denote the four-dimensional  and
$D$-dimensional Einstein tensors, respectively, while
  $a$ and $b$ are certain constants.
In order to be able to describe 4D gravity 
at short distances with the right value of the 
Newton's coupling we set
\beq
a\,+\,b\,=\,1\,.
\label{ab}
\eeq
Note, that the first two terms in parenthesis 
on the left-hand side of Eq.~(\ref {geq}) can be identically 
rewritten as 
\beq
&& (a+b) \, G^{(4)}_{\mu\nu}\,+\, 
b\left (-\partial_\mu \partial_a \, h^a_\nu \,- \,
\partial_\nu \partial_a \, h^a_\mu \,- \, \partial_a^2 h_{\mu\nu}\, + \,
\eta_{\mu\nu}\partial_a^2 h^C_C  \right.\nonumber\\[2mm]
&&+ \left.\partial_\mu \partial_\nu 
h^a_a \, -\,\eta_{\mu\nu} \partial_\alpha^2\,h^a_a\, + \, 2\eta_{\mu\nu}
\partial_a\partial_\alpha h^{a\alpha}\,+\,\eta_{\mu\nu}\partial_a\partial_b
h^{ab}\right) .
\eeq
The above  equation of motion (\ref {geq}) --
which should  be viewed as a regularized version of the DGP model --
could  be  obtained from the action   (\ref {1}) as well, provided
the latter is
amended by certain extrinsic curvature terms (for more details see
Ref.~\cite {AGV}). Below we will study this 
version of the regularized DGP model for  various values  
of the parameters $a$ and $b$.  Certain issues in the 
$a=0$, $b=1$  case, regularized with finite brane width, 
have been recently analyzed in Ref. \cite {MP}.
We will find below that phenomenologically more attractive 
is the  $a=1$, $b=0$ case.

The issues to be addressed are:

Assume that gravity measurements are done
at points $x_1$ and $x_2$ that are confined to the brane.
At distances $|x_1-x_2|< H_0^{-1}$ the graviton propagator 
{\it imitates} that of a massive 4D unstable particle with 
mass $\sim m_c$. Given that the model itself is intrinsically 
(4+$N$)-dimensional, the following  questions must be answered:

\vspace{2mm}

(i) Does the graviton propagator $G(x_1,\,\, x_2,\,\, y=0)$
satisfy the requirements of four-dimensional unitarity?

(ii) Do ``abnormalities"  occur at 4D momenta much smaller than
the ultraviolet (UV) cut-off and larger than the infrared (IR) crossover scale, 
such as a precocious onset of the strong-coupling regime?

\vspace{2mm}

Needless to say, the answers to these questions
determine whether the DGP model is intrinsically self-consistent
and phenomenologically viable.
The answer to the first question will be demonstrated to 
be {\em positive} while to the second {\em negative}. That is to say,
the  situation is {\em most favorable}.  We hasten to add that
it is not trivial to see that
this is indeed the case. It is necessary to carry
out a rather subtle analysis which circumvents
stereotypes in, at least,  one point. 

For what follows it is instructive to confront 
the DGP model with the 4D ``hard"-massive 
gravity \cite{PF} (or the Pauli-Fierz (PF)
gravity) which leads to a large-distance modification of
interactions too. (The action of the
``hard"-massive gravity  is given below in
Eq.~(\ref {PF}).) 
In particular, we compare the
perturbative treatments of these two models.
Perturbation theory in Newton's constant in the 4D massive
gravity breaks down at a scale much lower than the cut-off
scale of the theory --- this was first obtained for spherically
symmetric sources in Ref. \cite {Arkady}. The origin of this
breaking can be traced back \cite {DDGV} to  Feynman diagrams  involving
nonlinear interactions of gravitons. In terms of degrees of freedom,
it is the longitudinal polarizations of the massive gravitons that 
are responsible
for the perturbation theory breaking. This   can be readily inferred from
dynamics of these modes analyzed in Ref.~\cite {AGS}.
(Note also that the PF graviton propagates 
six degrees  of freedom instead of five \cite {BD} in the full quantum theory.
 This makes the corresponding Hamiltonian unbound from  below \cite {BD}. 
As a result,    solutions exist that destabilize the empty Minkowski 
space \cite {GG} and the instability can develop practically instantaneously.).

As will be shown,  in this respect
the DGP gravity (\ref{11})  presents a drastic improvement.
In contradistinction with the 4D Pauli-Fierz theory of massive gravity,
the precocious  breakdown of perturbation theory does not occur in
the DGP model  with $N\ge 2$. A direct analogy with the Higgs
mechanism for non-Abelian gauge fields
is in order here.  

For non-Abelian gauge fields with the hard
gauge-boson mass term, appropriate nonlinear amplitudes 
invalidate the perturbative
expansion (i.e. violate  the unitarity bound) at a scale set by the
gauge-boson mass divided by the gauge coupling constant.
To cure this disaster, one   introduces an extra scalar --- the Higgs
field.\,\footnote{The mass of the Higgs field  itself
needs  a stabilization mechanism.  This is a separate story, however.} 
By the same token, certain 
nonlinear perturbative amplitudes
of the 4D ``hard"-massive gravity  blow
up  precociously \cite {Arkady,DDGV,AGS}
at a scale significantly lower than a naive UV cut-off of
the theory under consideration.
The unwanted growth of the amplitudes is canceled, however, in
the DGP model  at the  expense of introducing an
infinite number of the KK fields. Thus, the action (\ref {11})
gives rise to a gravitational analog of the Higgs mechanism,
with an infinite number of  the ``Higgs'' fields which include both
vector and scalar states.

It has been  recently argued~\cite {DR} that
the spectrum of the DGP model  contains
tachyonic states with a negative norm (``tachyonic ghosts").
The conclusion was based on an analysis of the
poles in the graviton propagator derived from the action (\ref {1}).
In fact, the analysis of Ref. \cite {DR}  leaves aside subtle
points  of appropriately defining the Green's function poles.
We formulate and discuss  an appropriate  rule  for defining
the would-be poles. With this rule accepted, 
 the 4D unitarity of  the Green's functions
is guaranteed. This is not the case with respect to 4D analyticity and 
causality, though. In certain instances we will
have to sacrifice 4D causality at distances of the order of  
today's Hubble scale. As was argued in Ref. \cite {ADDG} this  
is a welcome feature for solving the cosmological constant problem.

The organization of the paper is as follows.
In Sect.~\ref{tsesf} we  discuss in detail a simplified
version of the phenomenon, with the tensorial structure suppressed.
We consider a scalar field with the Lagrangian similar to Eq.~(\ref{1})
and derive the Green's function. There are  complex poles of the Green's 
function on   the second and subsequent non-physical Riemann sheets.  
This corresponds to 
a resonance nature of the 4D massive scalar.
In Sect.~\ref{potg}  we discuss the same problem for
gravity, i.e. including the graviton tensorial
structure. The issue of proper definition of  poles in the graviton
propagator 
emerges in earnest in the trace part. In certain cases, 
a naive way of defining the poles leads to inconsistent results --- 
violations of unitarity. For these cases we point out a way out, 
treating the would-be poles in the
Green's function in such a way that  unitarity
is not violated.  Finally, in Sect.~\ref{ptimghvs}
 the $N\ge 2$ DGP model is argued to have no strong coupling problem
at intermediate scales, in contradistinction to the 4D Pauli-Fierz 
gravity with the hard mass term.

\section{The simplest example: scalar field}
\label{tsesf}

To warm up, we start our discussion with a simple model of
a scalar field $\Phi$ in $(4+N)$-dimensional space-time.
For convenience we separate
the dependence of the scalar field $\Phi$ on four-dimensional and
higher-dimensional coordinates,
$\Phi (x_\mu, y_a) \equiv \Phi(x,y)$.
The  two kinetic term action --- the scalar counterpart of
Eq.~(\ref {1}) --- has the form
\beq
S\,=\frac{\mpl^2}{2}\, \int d^4x\,\partial_\mu\Phi(x,0)\,
\partial^\mu\Phi(x,0)\,+\frac{\m^{2+N}}{2}\, \int d^4x\,d^Ny\,
\partial_A\Phi(x,y)\,\partial^A\Phi(x,y)\,.
\label{saction}
\eeq
It is important to understand that in
the scalar case the analog of the new term  
included in Eq.~(\ref {geq}) but absent in  (\ref {geq0})
reduces, identically, to the already  existing localized term.
This is a consequence of  our choice of the  regularization of 
the brane width $\Delta$ and  the boundary conditions
according to which  transverse derivatives vanish on the brane in the 
$\Delta \to 0$ limit. 

To study interactions mediated by the scalar field
we assume that $\Phi$  couples to a source $J$ localized in the
4D subspace in a conventional way, $\int d^4x \,  \Phi(x,0)\,  J(x)$.
Then the  equation  of motion takes the form
\beq
\d \, \mpl^2 \, \partial^2_\mu \,\Phi(x,0)\,+\,
\m^{2+N} \,\partial^2_A \, \Phi(x,y)\, = \, J(x)\, \d \,.
\label{seq}
\eeq
The very same equation applies to the scalar field Green's
function.

\subsection{Spectral  representation}
\label{dr}

At first, using the scalar field example,
we will summarize  general arguments for the 
existence/absence of a spectral representation in higher-dimension 
theories with the worldvolume kinetic terms as in Eq.~(\ref{saction}).  
Explicit formulae below 
refer to the scalar case. In the next section we will 
consider  gravity, with the appropriate tensorial
structure,  and will emphasize  crucial 
differences between the present scalar example  and full-blown gravity.

By the spectral representation we mean  the K\"allen-Lehmann (KL)
representation for the free tree-level propagator in the model 
(\ref {saction}) in terms of {\it four-dimensional} 
Mandelstam variables. Since the theory described 
by (\ref {saction}) is intrinsically higher-dimensional,
it is not clear {\it a priori} why the spectral representation 
in terms of the 4D variables should hold at all. Indeed, 
on the one hand,   the KL representation expresses
  the fact that a given amplitude, as a function of $p^2$
(the 4D momentum squared), is   {\it analytic}  in the complex
$p^2$ plane everywhere  except possible isolated poles   plus a
branch cut along the real positive  semiaxis. On the other hand, 
the model (\ref {saction}) is   nonlocal   from the 
4D standpoint; hence, it is not obvious why   analyticity of the 
amplitude with respect to the 4D variable  $p^2$ should take place
 in the regime 
where higher-dimensional effects become crucial.
  
There is an alternative point of view
on the KL representation of the model (\ref {saction}). 
We can assume that the extra dimensions $y$ are compactified, with
a finite (albeit arbitrarily large)
compactification radius $R$. Then, the spectrum of the theory 
must consist of ``discretized"  Kaluza-Klein modes. From 
the 4D standpoint they are just certain massive states. 
Then one  could certainly
obtain the KL representation by writing the (tree-level) 4D propagator 
summing up the entire tower  of the KK eigenstates.  

This strategy is readily implemented in 
the conventional compactifications, 
when the brane worldvolume term is absent. In this case
the spectrum of the Kaluza-Klein eigenmodes,
\beq
\Phi (x, y) = \sum_n \Phi_n (x) \phi_n (y)\,, 
\label{KKmodes}
\eeq
becomes trivially discrete, with the eigenvalues $m_n^2\sim n^2/R^2$.
If so, the expression for the Green's function
$G(p,y=0)$ takes the form
\beq
G(p,y=0) =\sum_n\,\,  \frac{\phi_n^* (0)\phi_n (0)}{m_n^2\, - \, 
p^2\, - \,i\e }\,,\qquad  p^2 =p_\mu p^\mu\,.
\label{aasr}
\eeq
In other words, $G(p,y=0)$ is the sum over the infinite number of
poles, with the positive-definite residues.
As $R\to\infty$ the sum goes into the standard dispersion integral,
\beq
G(p,y=0) =\frac{1}{\pi}\,\int_0^\infty\, dt\,  
\frac{\rho(t)}{t-p^2-i\e}\,,
\label{sspectral}
\eeq
where $\rho(t)$ is a positive-definite spectral density.

The argument above, as well as the simple 
representation (\ref{aasr}) or (\ref{sspectral})
following from it, neglects the existence of the 
brane worldvolume kinetic term in the action (the first term 
on the right-hand side of (\ref {saction})). This term is crucial,
and by no means can be neglected. It gives rise to kinetic mixings of the
KK modes on the brane worldvolume,
\beq
\frac{\mpl^2}{2}\, \int d^4x\,\left ( \sum_m \partial_\mu \Phi_m(x)\,
\phi_m(0) \right ) \left (\sum_n 
\partial^\mu \Phi_n(x)\, \phi_n(0) \right )\,.
\label{mixing}
\eeq
Therefore, the KK modes defined in Eq.~(\ref {KKmodes})
are {\em not} the  eigenstates of the Hamiltonian in the presence of 
the brane kinetic term. Diagonalization is needed. 
For the scalar field example (\ref {saction}),
explicit diagonalization is possible and was in fact carried out~\cite {DGKN}.
As a result, the spectral  representation can be argued to exist in the 
desired form, Eq.~(\ref {sspectral}).

In the case of gravity  things are more complicated, however.
The worldvolume EH term gives rise to kinetic mixings between
the massive KK modes of distinct spins. It is not 
obvious how to diagonalize the full linearized Hamiltonian.
Even if the diagonalization is possible  it is not clear whether 
the diagonal eigenstates are states of a definite 4D spin, 
and not the mixed states. This is all because of the large kinetic 
mixings between all the KK states on the brane.
Thus, the spectral representation we look for is hard (if possible at all) 
to obtain through explicit summation of the  eigenstates of the Hamiltonian. 
The best one can do is to write down the spectral representations in the
limiting  regimes when  the 4D EH term is either dominant 
or   negligibly small. We will return to this issue in the Sect.~\ref{potg}. 
Prior to delving in the gravity problem we want to conduct detailed studies 
of the scalar example (\ref {saction}).

\subsection{Solving Eq.~(\ref{seq}) in the general case}
\label{se6itgc}

To  solve this equation it is convenient to  Fourier-transform
it  with respect to ``our" four space-time coordinates
$x_\mu\to p_\mu$, keeping the extra $y$ coordinates intact. Marking 
the Fourier-transformed quantities by  tilde, 
\beq
\Phi(x,y)\, \to \, {\tilde \Phi}(p,y)\,,
\label{fourier}
\eeq
we then get from Eq.~(\ref{seq})
\beq
\d \, \mpl^2 (-p^2)\,{\tilde \Phi}(p,0)\,+\,
\m^{2+N} \,(-p^2 -  \Delta_y ) {\tilde \Phi}(p,y)\, =
\,{\tilde J}(p)\, \d \,,
\label{seqmom}
\eeq
where $p^2\equiv p_0^2-p_1^2-p_2^2-p_3^2$, and  the notation
\beq
\Delta_y \, \equiv \, \sum_{a=1}^N {\partial^2\over \partial y^2_a}
\label{Delta}
\eeq
is used.

We will look for the solution of Eq.~(\ref {seqmom}) in the 
following form:
\beq
{\tilde \Phi}(p,y)\,\equiv \, D(p,y)\,\chi(p) \,,
\label{decompos}
\eeq
where the function $D$ is defined as a solution of the equation
\beq
(- p^2 -  \Delta_y -i{\bar \e})\, D(p,y)\,= \, \d\,.
\label{D}
\eeq
Note that the function $D$ is uniquely determined only after 
 the $i{\bar \e}$ prescription  specified above is implemented.
We also introduce a convenient abbreviation
\beq
D_0(p)\, \equiv \,D(p, y=0)\,.
\label{D0}
\eeq

Now, it is quite obvious that a formal solution of Eq.~(\ref {seqmom}) 
can  be written 
in terms of the function $D$ as follows:
\beq
{\tilde \Phi}(p,y) \,=\,-\,{{\tilde J}(p)  \over \mpl^2}\,\,\,
{D(p, y) \over p^2\,D_0(p)\, - \,
\m^{2+N}/\mpl^2 }\,+c\,{\tilde \Phi}_{\rm hom}(p,y)\,,
\label{ssol}
\eeq
where ${\tilde \Phi}_{\rm hom}(p,y)$ is a general
solution of the corresponding  homogeneous equation
(i.e., Eq. (\ref {seqmom}) with the vanishing right-hand-side),
and $c$ is an arbitrary constant. Equation (\ref{ssol})
presents, in fact, the Green's function too, up to 
the factor $\tilde J(p)/ \mpl^2$, which must be
amputated. In particular, for the Green's function
on the brane we have
\beq
G(p,0)=\frac{\mpl^2}{\tilde J(p)}\,   {\tilde \Phi}(p,y=0) \,,
\qquad  G_{\rm hom}(p,0) =\frac{\mpl^2}{\tilde J(p)}\,
{\tilde \Phi}_{\rm hom}(p,y=0)\,,
\label{G}
\eeq
while for arbitrary values of $y$
\beq
G(p,y)\, = \,- \,{ D(p,y) \over p^2\, D_0(p) -u^N}\,
+\,c\,G_{\rm hom}(p,y)\, ,
\label{spoles}
\eeq
where
\beq
u^N \equiv \, \frac{\m^{2+N}}{\mpl^2}= m_c^2\, M_*^{N-2}\,.
\label{defu}
\eeq
The presence/absence of the homogeneous part is regulated by the $i\epsilon$
prescription.
Note that if the first term on the right-hand 
side of Eq.~(\ref {spoles})
has poles on the real axis of $p^2$, then the 
homogeneous equation has a solution
\beq
G_{\rm hom}(p,y)\,=\, D(p,y)\,\delta \left (p^2D_0(p) - 
u^N  \right )\,.
\label{homsc}
\eeq
This fact will play an essential role for gravity,
as   will be discussed in due course in  Sect.~\ref{potg}.

In what follows we will examine
the poles of  the Green's function $G(p,y)$.
The positions of these poles depend on
the functions $G_{\rm hom}(p,y)$, and
$D_0$ as defined in Eqs.~(\ref{D}) and  (\ref {D0}).
The choice of a particular rule of treatment of
the poles corresponds to the choice of appropriate 
boundary conditions in the coordinate space. Note that the latter are
dictated by physical constraints on the Green's function $G$ 
rather than on the auxiliary function $D$.

To get to the main point, we will try the 
simplest strategy of specifying the poles and check, 
{\em aposteriori}, whether  this strategy
is self-consistent. 
Let us put
$$c=0$$
 and define $D$ in the Euclidean momentum space.
Since in the Euclidean space the expression for $D$ is well-defined
and has no singularities,
\beq
D(p_E, q) =\frac{1}{p_E^2 +q^2}\,,\qquad 
D(p_E, q)\equiv \int d^N y \, e^{iqy}\, D(p_E, y)\,,
\label{deuclid}
\eeq
\vspace{2mm}
$$
q^2 = \sum_a (q^a )^2\,,
$$
one can perform  analytic continuation from the Euclidean space 
to Minkowski.  This is not the end of the story, however.
It is the Green's function $G$ that we are interested in, not the
auxiliary function $D$. As  will be explained  below,
the above  procedure is consistent, for
the following reason.  The function $G$ obtained in this way
has a cut extending from zero to infinity. 
In addition, we find two complex conjugate 
poles on the second {\em non-physical} Riemann sheet of the complex 
$p^2$ plane. Moreover, there are additional poles on subsequent unphysical 
sheets. 

Since the poles are not on the physical Riemann sheet,
they do not correspond to any asymptotic states of the 
theory. A pole  on the second Riemann sheet is a 
well-known signature of a resonance state \cite {Eden}. 
Therefore, our toy scalar ``gravity" is mediated by a massive resonance.
The resonance-mediated gravity was first discussed
in Refs.~\cite {GRS,CEH,DGP0} in a different brane-world model.

Before passing to  consideration of particular cases it is worth 
reminding that the Green's functions in the $N\ge 2$ DGP models 
need a UV regularization~\cite{DG,Wagner,Kiritsis,DHGS}. This 
has been already mentioned. An appropriate  UV regularization can be achieved 
either by introducing an explicit UV cutoff, or, alternatively, by 
keeping a non-zero brane width in a consistent manner 
(defined in Ref.~\cite{DHGS}). For brevity we choose the former 
prescription by consistently taking the limit of zero brane width.
However, we should stress that  all our  results hold equally  well 
in the brane-width regularization method of \cite{DHGS}.

\subsection{Six dimensions}
\label{sd}

It is instructive to demonstrate how things work by
considering  separately the six-dimensional case.
In six dimensions  sensitivity to the UV cutoff is only
logarithmic, and it is conceivable  that the results obtained in the
cut-off theory could be consistently matched  to those
of a more fundamental UV-completed theory-to-come.\footnote{
The $D> 6$ models of brane-induced gravity 
are power sensitive to  UV physics. In general one expects 
all sorts of higher derivative operators in this case.} 

It is not difficult to calculate
\beq
D_0(s) \, =\frac{1}{4\pi}\,  \ln  \left ( {\Lambda^2\over -s}\,+1\,\right )
\,,\qquad s\equiv p^2\,, 
\label{6D0}
\eeq
where $\Lambda^2$ is an ultra-violet cut-off. 
With this expression for $D_0$ the function $G(p^2, 0)$ develops a {\em cut}
on the positive semi-axes of $s$ due to the logarithmic behavior of
$D_0(s)$. This fact has a physical interpretation. Since the
extra dimensions are
non-compact in the model under consideration,
the spectrum of the theory, as seen from the 4D standpoint,
consists of an infinite gapless tower of the KK modes.  
This generates a
cut in the Green's function for $s$ ranging
from zero to $+\infty$.  

In addition, there might exist isolated
singular points  in $G(p^2,0)$. These singularities (for
$s\ll \Lambda^2$) are determined by the equation
\beq
G^{-1}(s,0)\equiv  s - m_c^2 \, \left[
\frac{1}{4\pi}\ln \left(\frac{\Lambda^2}{ -s}\right)\right]^{-1} =\,0 \,,
\label{spoles1}
\eeq
where
$m_c^2$ is defined in Eq.~(\ref{mc}).
Let us introduce the notation
\beq
s\, \equiv \, s_0\,{\rm exp} (i\gamma)\,,
\label{scom}
\eeq
where $s_0$ is a {\em real positive} number. Then,
Eq. (\ref {spoles1}) has two solutions of the form
\beq
s_{0}\approx 4\pi\,   m_c^2 \left[ \ln
\frac{\Lambda^2}{ m_c^2 }
\right]^{-1}\,,
\eeq
and 
\beq
\gamma_1 \, \simeq \,-\,{\pi \over {\rm log} (\Lambda^2/m_c^2) }\,
~~~\gamma_2 \, \simeq \, 2\pi \,+ \,{\pi \over {\rm log}
(\Lambda^2/m_c^2) }\,.
\label{mpole}
\eeq
We conclude that there are two complex-conjugate poles
on the nearby non-physical Riemann sheets. These poles
cannot be identified with any physical states
of the theory. They are, in fact, manifestations of a 
massive resonance state.  All other complex 
poles appear on subsequent nonphysical Riemann sheets.

\subsection{More than six dimensions}
\label{sevend}

Physics at $D>6$ is similar to that of the six-dimensional world which  
was described in Sect.~\ref{sd}. There are 
minor technical differences between odd- and even-dimensional
spaces, however, as we will discuss momentarily.

In seven dimensions we find
\beq
D_0(s) \,= \,{1\over 2\,\pi^2}\,\left\{ \Lambda \,-
\, \sqrt {-s}\,{\rm arctan}\left( {\Lambda \over \sqrt {-s} } 
\right)  \right\} \,.
\label{7D0}
\eeq
As in the 6D case, there is a branch cut. The cut in this case is due to
the dependence of the Green's function on $\sqrt{s}$.
No other singularities appear on the  physical Riemann sheet.
All  poles are on non-physical Riemann sheets, as previously.

In  the eight-dimensional space the 
expression for $D_0$ reads
\beq
D_0(s) \,= \, {1\over 16\,\pi^2}\,  \left\{       \Lambda^2 \,+\,s\,\left(
{\rm ln} \,  {\Lambda^2 \over -s}\,+ \, 1\,\right) \right\}\,.
\label{8D0}
\eeq
Again, we find a cut due to the logarithm,
 similar to that of the 6D case. All isolated singularities 
appear on non-physical Riemann sheets.

The nine-dimensional formula runs parallel to  that  in 
seven dimensions,
\beq
D_0(s)\,= \,{1\over 12\,\pi^3}\,\left\{ {\Lambda^3 \over 3} \,+\,s
\left( \Lambda \,-
\, \sqrt {-s}\,{\rm arctan} \,  {\Lambda \over \sqrt {-s} } 
 \right) \right\}\,.
\label{9D0}
\eeq
Finally, in ten dimensions 
\beq
D_0(s)\,= \,{1\over 128\,\pi^3}\, \left\{ {\Lambda^4 \over 2}
\, +\, s \left [ \,\Lambda^2 \,+\,s\,\left( 
{\rm ln} \, {\Lambda^2\over -s}\,+ \,1\,\right ) \right ] \right\}\,.
\label{10D0} 
\eeq
The pole structure of $G$ is identical to that of 
the eight-dimensional case.  Since the pattern is now 
well established and clear-cut,
there seems to be no need in  dwelling on higher dimensions. 

Before turning to gravitons we would like to make 
comments concerning the UV cutoff $\Lambda$. 
The crossover distance $r_c\sim m_c^{-1}$ 
depends on this scale:  in 6D the dependence is 
logarithmic, while  in $D>6$ this dependence presents a power-law
\cite {DG,Ignat}. Hence, the crossover scale 
in the $N\ge 2$ DGP models, unlike 
that in the $N=1$ model, is sensitive
to  particular details of the UV completion of the theory.  
Since in the present 
work we adopt an affective  low-energy field-theory strategy,
we are bound to follow the least favorable
scenario in which  the cutoff and the bulk gravity scale 
coincide with each other and both are equal to $\m \sim 10^{-3}$ eV. 
If a particular UV completion were available, it could well
happen that the UV cutoff and bulk gravity scale were different
from the above estimate. In fact, in 
the string-theory-based construction of Ref. ~\cite {Ignat} the 
UV completion is such
that  the cutoff and bulk gravity scale are in the ballpark of TeV.

In conclusion of this section it is
worth noting that the Green's function $D_0$
in the $N\ge 3$ case contains terms   responsible for 
branch cuts. These terms  are suppressed by  powers of $s/\Lambda $, 
and, naively, could have been neglected.
It is true, though,  that the explicit form of these terms is UV-sensitive
and cannot be established without the knowledge of UV physics.
One  should be aware of these terms since 
they  reflect   underlying physics --- the presence 
of the infinite tower of the KK states. Fortunately, none of the 
results of the present work depend on these terms.

\section{The graviton propagator}
\label{potg}

Now it is time to turn to  gravitons with their specific tensorial
structure. We will consider and analyze the equation of 
motion of the DGP-type model presented in Eq.~(\ref{geq}), which we 
reproduce here again for convenience
\beq
\d \, \mpl^2 \, \left ( a\,G^{(4)}_{\mu\nu}\,+\, b\,G^{(D)}_{\mu\nu} \right )
\delta^\mu_A\,\delta^\nu_B
\,+\, \m^{2+N} \,G^{(D)}_{AB}\,
= \, - T_{\mu\nu}(x)\, \delta^\mu_A\,\delta^\nu_B \, \d \,.
\label{geq1}
\eeq
Here   $G^{(4)}$ and $ G^{(D)}$ 
denote the four-dimensional  and
$D$-dimensional Einstein tensors, respectively,
while $a$ and $b$ are certain constants satisfying the constraint
 $$a+b=1\,.$$
For simplicity we choose    a source  term localized on the brane,
namely,  $T_{\mu\nu}(x)\d $. At the effective-theory level the ratio
$a/b \equiv a/(1-a)$ is a free parameter. The only guidelines
we have for its determination are (i) phenomenological viability; (ii)
intrinsic self-consistency of the effective theory
which, by assumption, emerges as a low-energy limit of a self-consistent 
UV-completed underlying ``prototheory." Specifying the
prototheory would allow one to fix the ratio $a/(1-a)$ in terms of fundamental
parameters.

\vspace{1mm}

Our task is to study the gravitational field produced by
the source $T_{\mu\nu}(x)\d $. To this end we 
linearize Eq.~(\ref{geq1}). If $g_{AB}\equiv \eta_{AB}+2h_{AB}$,
in the linearized in $h$ approximation
 we find
\beq
G^{(D)}_{AB}&=&\partial_D^2\,h_{AB}\,-\,\partial_A \,\partial_C\,
h^C_B \, -\,\partial_B \,\partial_C\,
h^C_A \nonumber \\[3mm]
&+& \partial_A\,\partial_B\,h^C_C \, -\, \eta_{AB}\,
\partial_D^2\,h^C_C \,+ \, \eta_{AB}\,\partial_C\,\partial_D\,h^{CD}\,,
\label{GD}
\eeq
where $\partial_D^2 \equiv \partial_D \partial^D$.
On the other hand, the four-dimensional Einstein tensor in the linearized
approximation is
\beq
G^{(4)}_{\mu\nu}&=& \partial_\beta^2\,h_{\mu\nu}\,-\,\partial_\mu \,
\partial_\alpha \,
h^\alpha_\nu \, -\,\partial_\nu  \,\partial_\alpha \,
h^\alpha_\mu \,+\,\partial_\mu\,\partial_\nu\,h^\alpha_\alpha
\,\nonumber \\[3mm]
 &-&  \eta_{\mu\nu}\,
\partial_\beta^2\,h^\alpha_\alpha \,+ \, \eta_{\mu\nu}\,\partial_\alpha
\,\partial_\beta\,h^{\alpha \beta}\,.
\label{G4}
\eeq
In what follows we will work in the harmonic gauge, 
\beq
\partial^A\,h_{AB}\,=\,{1\over 2}\,\partial_B\,h^C_C\,.
\label{har}
\eeq
The advantage of this gauge is that in this gauge the expression for $ G^{(D)}_{AB}$ significantly simplifies,
\beq
G^{(D)}_{AB}\,=\,\partial_D^2\,h_{AB}\,-\,{1\over 2}\,
\eta_{AB}\,\partial_D^2\,h^C_C \,.
\label{GDs}
\eeq
Additional conditions which are invoked to solve the
$\{ab\}$ and $\{a\mu\}$ components of the equations of motion are
\beq
h_{a\mu}=0,~~~~~h_{ab}\,=\,{1\over 2}\,\eta_{ab}\,h^C_C\,.
\label{abp}
\eeq
Using the last equation it is not difficult to obtain the relation
\beq
N\,h^\mu_\mu\,=\,(2-N)\,h^a_a \,.
\label{mua}
\eeq
This relation obviously suggests that we should consider separately 
two cases: 

(i)   $N=2$;

 (ii)   $N>2$.\\
 We will see, however, that
the results in the  $N=2$ and $N>2$ cases are somewhat similar.

\subsection{Brane-induced gravity in six dimensions (\boldmath{$N=2)$}}
\label{igisd}

In two extra dimensions  Eq. (\ref {mua}) implies
\beq
h^\mu_\mu\,=\,0\,.
\label{mumu0}
\eeq
Therefore, the trace of the $D$-dimensional graviton
coincides with the trace of the extra-dimensional part,
\beq
h^A_A\,=\,h^a_a\,.
\label{aaaa}
\eeq
As a result, the four-dimensional components
of the harmonic gauge condition (\ref {har})
reduce to
\beq
\partial^\mu \,h_{\mu\nu}\,=\,{1\over 2}\,\partial_\nu\,h^a_a\,.
\label{4dgauge}
\eeq
Let us now have a closer
look at the $\{\mu\nu \}$ part of Eq. (\ref {geq}).
Taking the trace of  this equation and using Eqs.~(\ref {GDs}), (\ref {G4}),
(\ref {mumu0}) and (\ref {4dgauge})
we arrive at\footnote{As before, we put the 
transverse derivatives to be zero in the $\Delta\to 0$ limit.}
\beq
(3b\,-\,1)\,\d \, \mpl^2 \, \partial^2_\mu \,h^a_a\,+\,
2\,\m^{2+N} \,\partial^2_A \, h^a_a\, = \, T^\mu_\mu\, \d \,.
\label{6Dtrace}
\eeq
The  obtained  equation is very similar to the scalar-field 
equation (\ref {seq}). Therefore, we will follow the same route as in  the scalar-field case,
until we come to a subtle point, a would-be obstacle,
which was non-existent  in the scalar-field case. 

Let us Fourier-transform  Eq. (\ref {6Dtrace}),
\beq
&&(3b\,-\,1) \,\d \, \mpl^2 (-p^2)\, {\tilde h}^a_a(p,y) 
\nonumber\\[3mm]
&& + 2
\m^{2+N} \,(-p^2 -  \Delta_y )\, {\tilde h}^a_a(p,y) =
{\tilde T}(p)\, \d \,.
\label{6Dtracemom}
\eeq
The general solution of the above equation is
\beq
&& {\tilde h}^a_a(p,y)=   {{\tilde T}(p)\over \mpl^2}\,\g (p,y) \,,
\label{defg}\\[4mm]
&& \g = { D(p,y) \over 2m_c^2 \,-\,(3b\,-\,1)  p^2\, D_0(p)\,}\,
+\,c\, \g_{\rm hom}\, ,
\label{g6D}
\eeq
where  the solution of the homogeneous equation takes the form
\beq
\g_{\rm hom}\,=\, D(p,y)\,\delta \left (2m_c^2 \,-\, (3b\,-\,1) p^2D_0(p) 
\right )\,.
\label{hom6D}
\eeq
To begin with, let us consider  the case  $3b>1$. Then
the first term on the right-hand side  of Eq.~(\ref {g6D}) has  poles
for complex values of  $p^2$,  as can be  readily seen from the expressions for
$D_0$ obtained in  Sect.~\ref{tsesf}.  For instance, in the 6D case
this pole is determined by the equation
\beq
s= {2m_c^2\over (3 b - 1) D_0(s)}= {4\pi \,{2m_c^2}\over (3b-1)}\, 
\left[\ln \frac{\Lambda^2}{-s} \right]^{-1}\,.
\label{6Dnewpoles}
\eeq
This equation has at least two solutions of the form
\beq
s_{*}\approx {4\pi\,   2m_c^2 \over 3b-1}
\left[ \ln \frac{\Lambda^2}{ m_c^2 }
\right]^{-1}\,,
\label{som}
\eeq
and 
\beq
\gamma_1 \, \simeq \,-\,{\pi \over {\rm log} (\Lambda^2/m_c^2) }\,
~~~\gamma_2 \, \simeq \, 2\pi \,+ \,{\pi \over {\rm log}
(\Lambda^2/m_c^2) }\,.
\label{*pole}
\eeq
The quantity ${\tilde h}^a_a(p,y) $ is not a gauge invariant variable. 
Therefore, the presence  of certain poles in the expression for
${\tilde h}^a_a(p,y)$ depends on a gauge. However, explicit 
calculations (see below) show that the poles found above 
also enter the gauge invariant physical amplitude. 
Therefore, we need to take these poles seriously and analyze their
physical consequences.

\subsection{$b>1/3$}
\label{2bba}

If $b>1/3$  there are no poles on the physical Riemann sheet.
Instead,   poles appear on the nearest non-physical 
Riemann sheets. These poles cannot be identified with any physical states
of the theory. They represent a signature of  
massive resonance states.  All other complex  poles appear on subsequent nonphysical Riemann sheets.

Using a contour integral one can easily write down the spectral representation 
for the Green's function ${\cal G}$
\beq
{\g}(p,y=0) \,= \, {1\over \pi}\,\int_0^{\infty}\,
{ \rho (t) \,dt \over \,t\,- \,p^2\,-\,i\,{\bar \e}}\,,
\label{disp4D}
\eeq 
where the spectral function is defined as 
\beq
\rho (t)\,=\,{2\,m_c^2\, {\rm Im}\, D_0(t) 
\over\left[ (3b-1) t\,{\rm Re}\, D_0\,-\,2m_c^2\right]^2\,+\left[(3b-1)t\,
{\rm Im}\, D_0)\right]^2}\,,
\label{rho}
\eeq 
and  
\beq
{\rm Im}D_0\,=\,\pi\,\int \,{d^N q\over (2\pi)^N}\,\delta(t-q^2)\,
=\,{\pi^{N+2\over 2} \over (2\pi)^N \Gamma(N/2)}\,
t^{N-2\over 2}\,.
\label{ImD0}
\eeq 
We see that $\rho(t)$ satisfies the positivity requirement.
Equation (\ref {disp4D}) guarantees that the Green's 
function $ \g $ is causal.

The next  step is applying the expression 
for $\g$ to   calculate ${\tilde h}_{\mu\nu}$. In fact, it is more convenient 
to calculate the tree-level amplitude
\beq
A(p,y)\,\equiv \,{\tilde h}_{\mu\nu}(p,y)\,{\tilde T}^{\prime\,\mu\nu}
(p)\,,
\label{amp}
\eeq
where ${\tilde T}^{\prime\,\mu\nu}(p)$ is a
conserved energy momentum tensor,
 $$
 p_\mu\,{\tilde T}^{\prime\,\mu\nu}\,
=\,p_\nu\,{\tilde T}^{\prime\,\mu\nu}\,=\,0\,.
$$
Using Eqs. (\ref{geq1}),  (\ref {defg}) and  (\ref{6Dc}) 
we obtain the following expression for the amplitude $A(p,y)$:
\beq
A(p,y)\,=\,{1\over \mpl^2}\,{D(p,y)\over p^2\,D_0(p)\,-\,m_c^2}\,
\left \{{\tilde T}_{\mu\nu}  {\tilde T}^{\prime\,\mu\nu}\,-\,
{{\tilde T}\, {\tilde T}^{\prime}   \over 2}\,
\left[ {(2b-1)p^2D_0 -m_c^2   \over (3b-1)p^2D_0 - 2m_c^2 }  
  \right] \right \} .
\label{a}
\eeq

\vspace{2mm}

\noindent
Let us study  the above expression in some detail. The first
question to ask is about poles.
It is quite clear that the $p^2$-poles of
 $A$ are of two types; their position is determined by:
$$
p^2\,D_0(p)\,=\,m_c^2
$$ 
or
$$
(3b-1) p^2\,D_0(p)\,=\,2m_c^2\,.
$$
 As was explained previously, 
all these poles appear on the second  Riemann sheet,
 with the additional  images on other non-physical sheets.
None of these poles can be identified with asymptotic physical 
states. As was elucidated above, the occurrence
of the poles on the second and subsequent Riemann sheets corresponds
to the massive-resonance nature of the effective 4D graviton.
Our previous analysis can be repeated practically {\em verbatim},
 with minor modifications, 
proving analyticity and causality of the amplitude $A$.

Next, we observe that at  large momenta, i.e., 
when $p^2\,D_0(p)\gg m_c^2$, the scalar part of 
the propagator has 4D behavior;  the tensorial structure 
is not four-dimensional, however. 
The   terms in the   braces  in Eq.~(\ref {a}), namely, 
\beq
\left \{{\tilde T}_{\mu\nu}  {\tilde T}^{\prime\,\mu\nu}\,-\,
{2b-1\over 2(3b-1)}\, {\tilde T}\, {\tilde T}^{\prime}\,\right \}\,,
\label{2polar}
\eeq
correspond to the exchange of massive gravitons and
scalar degrees of freedom. This would give rise to additional 
contributions in the light bending, and is excluded phenomenologically, 
unless the contribution  due to  extra polarizations is canceled 
by some other interactions (such as, e.g.,  an additional  
repulsive vector exchange). Note also that when $b\gg a$, i.e., 
$b\to 1$, one obtains the 
tensorial structure of  6D gravity, as expected from (\ref {11}).

On the other hand, at large distances,
i.e., at $p^2\,D_0(p)\ll m_c^2$, we get 
the following tensorial structure of the amplitude (\ref {a}): 
\beq
\left \{{\tilde T}_{\mu\nu}  {\tilde T}^{\prime\,\mu\nu}\,-\,
{1\over 4}\, {\tilde T}\, {\tilde T}^{\prime}\,\right \}\,.
\label{6Dpolar}
\eeq
This exactly  corresponds to the exchange of a six-dimensional 
graviton, as was expected.

\subsection{$b<1/3$}
\label{2bma}

This case is conceptually different from that of Sect.~\ref{2bba}.
As we will see momentarily, if $b<1/3$ there are no
problems in (i) maintaining 4D unitarity; and (ii)
getting the appropriate 4D  tensorial structure of gravity
at sub-horizon distances. This is achieved at a price of
abandoning 4D analyticity, in its standard form,
which could presumably lead to the loss of causality
at distances of the order of $m_c^{-1}\sim 10^{28}$ cm. 
The absence of  causality at distances   $ \gsim 10^{28}$ cm,
was argued recently \cite{ADDG} to be an essential 
ingredient for solving the cosmological constant problem.

Although all derivations and conclusions are quite similar for
any ratio $a/b$ as long as $2b<a$, we will stick to the technically
simplest example  $b=0$, $a=1$.
In the situation at hand,  the homogeneous 
part (\ref {hom6D}) need not
be trivial, i.e $c$ need not vanish.
The value of the constant $c$ is  determined once 
the rules for the pole at $ p^2D_0(p) + 2m_c^2 =0 $ are specified.
In Ref.~\cite {DR} the vanishing of $c$  was postulated. This choice leads
to  {\em non-unitary} Green's function.
Therefore, we abandon the condition  $c=0$ in an attempt to make
a more consistent choice that would guarantee 4D unitarity.
We stress that we are after unitarity here, not unitarity plus causality.

To begin with we pass to the Euclidean space in $p^2$ (i.e. $p^2\to p_E^2$)
and introduce the following notation:
\beq
P^{(E)}(p_E^2)\,\equiv\, {1\over 2\,m^2_c\,-\,p_E^2\,D_0(p_E)-\,i\e}\,.
\label{PE}
\eeq
The function $P^{(E)}$    
is  a Euclidean-space solution of Eq. (\ref {6Dtracemom}), 
with the particular choice $c=i\pi$. (The choice 
$c=-i\pi$ would lead to Eq.~(\ref {PE}) with the 
replacement  $\e \to -\e$). 

As the next  step we will analyze  the complex plane of $p_E^2$.
Since the function $D_0(p_E)$ is real, the function
$P^{(E)}(p_E^2)$ must (and does) have an isolated singularity in the $p_E^2$ plane
 which is similar to a conventional massive pole, except that it lies
 in the Euclidean domain. 
This singularity occurs at the point $p_E^2=p_*^2$  
is defined by the condition
\beq
p_*^2D_0(p_*)\,=\, 2\,m_c^2\,,~~~~~p_*^2\,\,\,\mbox{real and positive}. 
\label{pstar}
\eeq
This is the only isolated singularity in Eq.~(\ref {PE});
it is located in the complex  $p_E^2$ plane 
on the real positive semiaxis. In addition to this pole
singularity, the 
function (\ref {PE}) has a branch cut stretching from zero to
$-\infty$ due to the imaginary part of $D_0(p_E^2)$ 
appearing at negative values of $p_E^2$.
As before, this branch cut is the reflection
of an infinite gapless tower of the KK states.  
As a result, the following spectral representation obviously emerges
 for $P^{(E)}(p_E^2)$:
\beq
P^{(E)}(p_E^2+i\e)\,=\, 
{1\over \pi}\,\int_0^{-\infty}\,
{{\rm Im}P^{(E)}(u)\,du \over u\,-\,p_E^2\,}\,
+\, {R\over p_*^2 \,-\,p_E^2 \,-\,i\e}\,,
\label{PEdisp}
\eeq
with the Euclidean pole term being ``unconventional."
The residue of the pole $R$ is given (for any $N$) by
\beq
R^{-1}\,=\,\int\, {d^Nq\over (2\pi)^N}\,{q^2\over (q^2\,+\, p_*^2)^2}\,.
\label{R}
\eeq

Note that in the first term on the right-hand side 
of Eq.~(\ref {PEdisp}) the integration  runs from zero to minus infinity; thus,
the integrand never hits the would-be pole at $u=p_E^2>0$. 
Therefore, the $i\e$ prescription is in fact used only to specify the 
isolated pole at $p_E^2=p_*^2$.

We proceed further and define a {\it symmetric} function
\beq
\Pi^{(E)}(p_E^2)\, \equiv \, { 1 \over 2} \left\{
P^{(E)}(p_E^2-i\e)\,+\,P^{(E)}(p_E^2+i\e) \right\}\,.
\label{PiE}
\eeq
It is just this symmetric function on which we will focus
in the remainder of the section.
Let us return to Minkowski space. This is done by substituting
$$
 p_E^2 \to {\rm exp}(-i\pi)p^2\,,\qquad  u\to {\rm exp}(-i\pi) t
 $$
in Eq.~(\ref {PEdisp}). Furthermore, observing that
${\rm Im}\, P ={\rm Im}\, \Pi $, we obtain the following
representation for the Minkowskian $\Pi$:
\beq
\Pi (p) \,=\,{1\over \pi}\,\int_0^{\infty}\,
{{\rm Im}\Pi (t)\,dt \over t\,-\,p^2\,-\,i\,{\bar \e}}\, +\,\Pi_0(p)\,,
\label{Pi}
\eeq
where 
\beq
\Pi_0(p)\,\equiv \,{1\over 2}   \left ( 
{R\over p_*^2 \,+p^2 \,-\,i\e}\,+\,
{R\over p_*^2 \,+p^2 \,+\,i\e} \right )\,.
\label{Pi0}
\eeq
It is necessary to emphasize
 that $\epsilon $ and $\bar\epsilon $ are two distinct regularizing
parameters, $\epsilon\neq\bar\epsilon$. The parameter $\epsilon$ is used 
to regularize the pole at $p^2=-p_*^2$, while $\bar\epsilon$
sets the rules for the branch cut.
The most important property of $\Pi$    is 
that the pole at $p^2=-p_*^2$ has no imaginary  part, by construction. Hence, 
there is no physical particle that corresponds to this pole.  
In the conventional local field theory the only 
possible additions with no imaginary part are polynomials.
Here we encounter a new structure which will be discussed in more detail
at the end of this section. 

Our goal   is to show that a 4D-unitarity-compliant  spectral representation holds
for the Green's function on the brane, at least in the domain where 
the laws of 4D physics are applicable. To this end we turn to the 
function $\g(p,y)$,   defined as  
\beq
\g \,=\, {D(p,y)}\,\Pi(p^2)\,.
\label{6Dc}
\eeq
with the purpose of studying its properties. 
It is convenient to pass to the momentum space with respect to 
extra coordinates too. Then, the propagator (\ref {6Dc}) 
takes the form
\beq
{\tilde \g}(p,q)\,=\,{\Pi(p^2)\over q^2 \,-\, p^2\, - i\,{\bar \e}}\,.
\label{pq}
\eeq
With these definitions in hand, we can write down 
the 4D  dispersion relation. We start from  the 
K\"allen-Lehman representation for the propagator (\ref {pq}).
As we will check below, this representation takes the form 
\beq
{\tilde \g}(p,q)\,= \, {1\over \pi}\,\int_0^{\infty}\,
{{\rm Im}{\tilde \g}(t,q) \,dt \over t\,-\,p^2\,-\,i\,{\bar \e}}\,
+\,{\Pi_0(p^2) \,-\,\Pi_0(q^2) \over q^2\, -\,p^2\,- 
\,i\,{\bar \e}}\,.
\label{disp}
\eeq
The imaginary part of ${\tilde \g}$ is defined as follows
\beq
{\rm Im}{\tilde \g}(t, q)\,=\,\pi \,\delta (q^2\,-\,t)\, 
{\rm Re}\Pi(t)\,+\,
{\rm Im}\Pi(t)\,{\cal P}{1\over q^2\,-\,t}\,,
\label{Im}
\eeq
where ${\cal P}$ stands for the {\it principle value} of a singular function,
\beq
{\cal P} \, {1\over q^2\,-\,t}\, = \,{1\over 2}
\left ( {1\over q^2\,-\,t\,+\,i\delta} \, + \,
 {1\over q^2\,-\,t\,-\,i\delta}\right ). 
\label{P}
\eeq
The fact that Eq. (\ref {disp}) holds can be checked by substituting 
(\ref {P}) and (\ref {Im}) into (\ref {disp}) 
and exploiting the relation
\beq
{1\over \pi}\,\int_0^{\infty}\,
{{\rm Im}{\Pi}(t) \over t\,-\,p^2 \,-\,i\,{\bar \e}}\,
{\cal P}{1\over q^2\,-\,t }\,dt \,=\,- 
{{\rm Re}\Pi(q^2) \,-\,\Pi_0(q^2)\,+\,\Pi_0(p^2)\,-\,\Pi(p^2)
\over  q^2-p^2 - i\,{\bar \e}}\,.
\label{check1}
\eeq
This turns Eq. (\ref {disp}) into  identity.

Finally we approach  the main point  of this  section -- 
the dispersion relation for $\g (p,y=0)$,
 the Green's function on the brane.
As such, it must
  have a spectral representation with the positive 
spectral density, as we have already seen from the KK-based analysis. 
The positivity is in one-to-one correspondence with the 4D unitarity.

The dispersion relation can be obtained by integrating
(\ref {disp}) with respect to $q$,
\beq
{\g}(p,y=0)\,= \, {1 \over \pi}\,\int_0^{\infty}\,
{ \rho (t) \,dt \over \,t\,- \,p^2\,-\,i\,{\bar \e}}\,
+\,\Pi_0(p^2)\,{\rm Re}D_0(-p_*^2)\,.
\label{disp4Dp}
\eeq 
According to Eq. (\ref {disp}), the spectral density 
$\rho$ is defined as
\beq
\rho (t)\,=\,\int \,{d^N q\over (2\pi)^N}\,
{\rm Im}\, {\tilde \g}(t, q)\,.
\label{rhoDEF}
\eeq 
The first term on the right-hand side
in Eq.~(\ref{disp4Dp}) is conventional while the second is not,
and we hasten to discuss it. This term has no imaginary part, by construction.
Hence, it does not contribute to the unitarity cuts
in diagrams. Therefore, this term does not affect the 
spectral properties. 

As was mentioned, in conventional 4D theories 
only a finite-order polynomial in 
$p^2$ that has no imaginary part can be added to or subtracted from 
the dispersion relation. This is because, normally one  deals
 with Lagrangians which contain only a finite 
number of derivatives, i.e., a finite number of terms  
with positive powers of  $p^2$ in the momentum space.
In the problem under consideration  this is not the case, however.
In fact, no  local 4D Lagrangian exists in our model 
 at all,  and yet we are studying the spectral 
properties in terms of the intrinsically 4D variable,  $p^2$. 
The theory (\ref {1}) is inherently higher-dimensional because
of the infinite volume of  the extra space. One can try to
``squeeze" it in four dimensions at a price of having 
 an  {\it infinite} number of 4D fields.  
For such a theory  there is no  guarantee that  
{\it analyticity} of the Green's functions 
in terms of the 4D variable $p^2$ will hold because 
 the  effective  4D Lagrangian
obtained by ``integrating out'' the infinite gapless 
KK tower  will necessarily contain \cite{DGS} non-local terms of the
type  $\partial^{-2}$. (Note that a similar prescription  for the poles 
in a pure 4D local theory \cite {Lee} is hard to reconcile with 
the path integral formulation \cite {Gross}. In our  case this is not 
a  concern  since the theory is not local in four-dimensions 
in the first place.)

Therefore, it is only natural that 4D unitarity can be maintained but 4D 
analyticity cannot. Non-analyticity 
leads to violation of causality, generally speaking. That is to say, the 
Green's function (\ref {disp4Dp}) is acausal. 
Therefore, we have an apparent violation of causality 
in the 4D slice of the entire $(4+N)$ dimensional theory
which, by itself, {\em is causal}. The apparent acausal effects can 
  manifest  themselves only at the  scale  of the order of
$m^{-1}_c\sim 10^{28}$ cm. In fact, as was noted in \cite{ADDG},
this is a welcome feature  for a possible solution of the cosmological 
constant problem.

Let us now return to the first term on the right-hand side
of Eqs. (\ref {disp4Dp}). Using Eqs.~(\ref {Pi}) and (\ref {Im}) we can calculate
the spectral function which comes out as follows:
\beq
\rho (t)\,=\,{2\,m_c^2\, {\rm Im}D_0(t)\over (t\,{\rm Re}D_0\,+
\,2m_c^2)^2\,+\,(t\,{\rm Im}D_0)^2}\,,
\label{rhop}
\eeq 
where 
\beq
{\rm Im}D_0\,=\,\pi\,\int \,{d^N q\over (2\pi)^N}\,\delta(t-q^2)\,
=\,{\pi^{N+2\over 2} \over (2\pi)^N \Gamma(N/2)}\,
t^{N-2\over 2}\,.
\label{ImD0p}
\eeq 
We see that $\rho(t)$ satisfies the positivity 
requirement.\,\footnote{For $N\ge 5$ 
the integral in Eq.~(\ref {disp4Dp}) diverges. However, since 
our model has a manifest UV cutoff $\Lambda$, the above integral 
must be cut off at $\Lambda$. Alternatively, one could use
a dispersion relation with subtractions.}
 
\vspace{1mm}

Next we observe that at  large momenta, i.e., 
at $p^2\,D_0(p)\gg m_c^2$, the propagator we got
has the desired 4D behavior. For the scalar part of the propagator this 
is expected from the studies of  Sect.~\ref{tsesf}.
However, with regards to the tensorial structure 
this circumstance is less trivial.
If $p^2\,D_0(p)\gg m_c^2$ the   terms in the  braces in Eq.~(\ref {a}), 
\beq
\left \{{\tilde T}_{\mu\nu}  {\tilde T}^{\prime\,\mu\nu}\,-\,
{1\over 2}\, {\tilde T}\, {\tilde T}^{\prime}\,\right \}\,,
\label{2polarp}
\eeq
correspond to the exchange of two physical graviton polarizations.
 Therefore, for the observable 
distances  the tensorial structure of the massless
4D graviton (\ref {2polarp}) is recovered. 

On the other hand, for large  (super-horizon) distances,
 $p^2\,D_0(p)\ll m_c^2$, we get 
a different tensorial structure of the same  amplitude,
\beq
\left \{{\tilde T}_{\mu\nu}  {\tilde T}^{\prime\,\mu\nu}\,-\,
{1\over 4}\, {\tilde T}\, {\tilde T}^{\prime}\,\right \}\,.
\label{6Dpolarp}
\eeq
This exactly corresponds to the exchange of the six-dimensional 
graviton.

\subsection{\boldmath{$D>6$}}
\label{dls}

Corresponding calculations and results are quite similar to the
$D=6$ case, with minor technical distinctions which we summarize below.
For $N\neq 2$
\beq
h_{ab}\,=\,{1\over 2\,-\,N}\,\eta_{ab}\,h^\mu_\mu\,.
\label{hab}
\eeq
Therefore,  we get
\beq
\partial^\mu \,h_{\mu\nu}\,=\,{1\over 2\,-\,N}\,\partial_\nu\,
h^\alpha_\alpha\,.
\label{hmn}
\eeq
Then, the trace of the $\{\mu\nu\}$ equation takes the form
\beq
&& k_N\, \d \, \mpl^2 \,
  \partial^2_\mu \,h^a_a\,+\,
\m^{2+N} \,(N+2)\, \partial^2_A \, h^a_a
\nonumber\\[3mm]
 =&& {N} \, T^\nu_\nu\, \d \,.
\label{traceeqD}
\eeq
where $k_N\equiv 2-N(2-3b)$. The above equation 
which  can be used to find the solution we are after. 
We proceed parallel to  the 
six-dimensional case.  Let us introduce the notation
\beq
{\tilde h}^a_a(p,y)\, = \,{N}\, 
{{\tilde T}(p)\over \mpl^2}\,\g_N (p,y) \,,
\label{defgN}
\eeq
where 
\beq
\g_N = { D(p,y) \over - k_N\,p^2\,D_0(p) +  u^N\,(N+2)}
+\,c\, \g_{\rm N\,hom}\, .
\label{gD}
\eeq
The solution of the homogeneous equation takes the form
\beq
\g_{\rm N\, hom}\,=\, D(p,y)\,\delta \left (-k_N\,p^2\,D_0(p) \,+\, u^N
\,(N+2) \right )\,.
\label{homD}
\eeq
Here  $$u^N\equiv \m^{2+N}/\mpl^2\,.$$
As in the 6D case, we conclude that that there exists a 
solution to the equation
$$
-k_N p^2\,D_0(p) \,+\, u^N(N+2)  =0
$$  
with a 
complex value of $p^2$. These poles occurs on the nonphysical sheets
as long as $k_N>0$, so the Green's function admits the spectral representation.

Using the expressions above one readily calculates  the tree-level amplitude $A$,
\beq
A(p,y)&=& {1\over \mpl^2}\,{D(p,y)\over p^2\,D_0(p)\,-\,
u^N }\nonumber \\[3mm]
&\times & 
\left \{{\tilde T}_{\mu\nu}  {\tilde T}^{\prime\,\mu\nu}\,-\,
{ {\tilde T}\, {\tilde T}^{\prime}  \over 2}\,
\left ( {(k_N-bN)p^2 D_0 - 2u^N  \over k_N p^2 D_0 - (2+N)u^N}    \right )
\right \}\,.
\label{aD}
\eeq

\subsection{$b> (2N-2)/3N$}
\label{2bbap}

In this case there are no poles on the physical Riemann sheet.
Hence, all the poles are of the resonance type. The tensorial structure 
at large distances is that  of the D-dimensional theory
\beq
\left \{{\tilde T}_{\mu\nu}  {\tilde T}^{\prime\,\mu\nu}\,-\,
{1\over 2+N}\, {\tilde T}\, {\tilde T}^{\prime}\,\right \}\,.
\label{DpolarD}
\eeq 
However, the tensorial structure at short distances $\lsim m_c^{-1}$
differs from that of 4D massless gravity. Hence, some additional 
interactions, e.g., repulsion due to a vector field, 
is needed to make this theory consistent with   data.

\subsection{$b< (2N-2)/3N$}
\label{2bmap}

The consideration below is very similar to the 6D case.
In perfect  parallel with the 6D case we consider for simplicity 
only the $b=0$ case and define the function
\beq
P_N^{(E)}(p_E^2)\,\equiv\, {1\over u^N(N+2)/(2N-2) \,-\,
 p_E^2\,D_0(p_E)-\,i\e}\,,
\label{PEN}
\eeq
which  has a spectral representation:
\beq
P_N^{(E)}(p_E^2+i\e)\,=\, 
{1 \over \pi}\,\int_0^{-\infty}\,
{{\rm Im}P_N^{(E)}(u)\,du \over u\,-\,p_E^2\,}\,
+\, {R\over p_*^2 \,-\,p_E^2 \,-\,i\e}\,.
\label{PENdisp}
\eeq
The residue $R$ is determined by Eq. (\ref {R})
while  $p_*^2$   is now  a solution to the equation
\beq
p_*^2D_0(p_*)\,=\, {u^N (N+2)\over 2(N-1)}\,,~~~~~p_*^2>0\,. 
\label{pstarN}
\eeq
As in the 6D situation, we use the expression (\ref {PENdisp}) to define 
a {\it symmetric} function 
\beq
\Pi^{(E)}_N(p_E^2)\, \equiv \, { 1 \over 2} \left\{
P^{(E)}_N(p_E^2-i\e)\,+\,P^{(E)}_N(p_E^2+i\e) \right\}\,.
\label{PiNE}
\eeq
The latter, being continued to the Minkowski space, admits the following 
spectral representation:
\beq
\Pi_N (p) \,=\,{1\over \pi}\,\int_0^{\infty}\,
{{\rm Im}\Pi (t)\,dt \over t\,-\,p^2\,-\,i\,{\bar \e}}\, +\,\Pi^{(N)}_0(p)\,,
\label{PiN}
\eeq
where 
\beq
\Pi^{(N)}_0(p)\,\equiv \,{1\over 2}   \left ( 
{R\over p_*^2 \,+p^2 \,-\,i\e}\,+\,
{R\over p_*^2 \,+p^2 \,+\,i\e} \right )\,.
\label{Pi0N}
\eeq
As previously,  $\epsilon $ and $\bar\epsilon $ are 
two {\em distinct} regularizing parameters, $\epsilon\neq\bar\epsilon$.

For the Green's function of interest  
\beq
\g_N \,=\, {{D(p,y)}\,\Pi_N(p^2)\over 2N-2} 
\label{Dc}
\eeq
we repeat  the analysis of Sects.~\ref{igisd},  \ref{2bba} and \ref{2bma}  to confirm
with certainty that the function $\g_N(p,y=0)$ does admit the spectral representation
(\ref {disp4Dp}), with a positive spectral function, 
similar to the 6D case, see Eq.~ (\ref {rhop}). 

The  expression in Eq.~(\ref {aD}) interpolates between
the four-dimensional and D-dimensional patterns.
This was already established for the scalar 
part of the amplitude in Sect.~\ref{tsesf}. Let  us examine  the 
tensorial part. For $p^2\,D_0(p)\gg u^N$ we get 
\beq
\left \{{\tilde T}_{\mu\nu}  {\tilde T}^{\prime\,\mu\nu}\,-\,
{1\over 2}\, {\tilde T}\, {\tilde T}^{\prime}\,\right \}\,.
\label{4DpolarD}
\eeq
This corresponds to two helicities of the 4D massless graviton.
In the opposite limit, $p^2\,D_0(p)\ll u^N$, we recover 
the tensorial structure corresponding to the $(4+N)$-dimensional 
massless graviton.

\section{Perturbation theory in massive gravity: \\
 hard mass vs. soft }
\label{ptimghvs}

We start from a brief review of the well-known
phenomenon --- the breakdown of perturbation theory for
the graviton with  the hard mass \cite {PF}, occurring at the
 scale   lower that the UV cutoff of the theory
\cite {Arkady,DDGV,AGS}. We then elucidate as to how
this problem is avoided in the models (\ref {1}, \ref{11}).

The 4D action of a massive graviton is
\beq
S_m\,=\,{\mpl^2\over 2} \,\int\,d^4x\,\sqrt{g}\,R(g)\,+ 
\,{\mpl^2\,m_g^2 \over 2}\,
\int\,d^4x\,\left (h_{\mu\nu}^2\,-\,(h^\mu_\mu)^2\right )\,,
\label{PF}
\eeq
where $m_g$ stands for the graviton mass and $h_{\mu\nu}\equiv
(g_{\mu\nu} -\eta_{\mu\nu})/2$. The mass term has the Pauli-Fierz form
  \cite {PF}. This is the only Lorentz invariant form of the
mass term which in the quadratic order in $h_{\mu\nu}$
does not  give rise to ghosts \cite {Neu}.
Higher powers in  $h$  could be  arbitrarily
added to the mass term  since there is
no principle, such as reparametrization
invariance, which could fix these terms.
Hence, for definiteness, we assume that
the  indices in the mass term are raised and lowered by
$\eta_{\mu\nu}$. Had we used  $g_{\mu\nu}$ instead,
the difference would appear only in   cubic and higher
orders in $h$,  which are not fixed anyway.

In order to reveal  the origin of the problem
let us have a closer look at the  free graviton
propagators in the massless and massive theory.
For the massless graviton we find
\beq
D^0_{\mu\nu ;\alpha\beta}(p)\,=\, \left(
{1\over 2} \,{\bar \eta}_{\mu\alpha}  {\bar \eta}_{\nu\beta}+
{1\over 2} \, {\bar \eta}_{\mu\beta}  {\bar \eta}_{\nu\alpha}-
{1\over 2} \, {\bar \eta}_{\mu\nu}  {\bar \eta}_
{\alpha\beta}\right)\frac{1}{
-\,p^2\,-\,i\epsilon}\,,
\label{4D}
\eeq
where
\begin{equation}
\bar\eta_{\mu\nu}\,\equiv \,\eta_{\mu\nu}\,-\,
\frac{p_\mu p_\nu}{p^2}\,.
\label{singp}
\end{equation}
The momentum dependent parts of the tensor structure were
chosen in a particular gauge   convenient for our discussion.
On the other hand, there is no gauge freedom for the
massive gravity presented  by the action (\ref{PF}); hence the
corresponding propagator is unambiguously determined,
\beq
D^m_{\mu\nu ;\alpha\beta}(p)\,=\,
\left(
{1\over 2} \,\tilde\eta_{\mu\alpha} \tilde \eta_{\nu\beta}+
{1\over 2} \, \tilde\eta_{\mu\beta}  \tilde\eta_{\nu\alpha}-
{1\over 3} \, \tilde\eta_{\mu\nu} \tilde 
\eta_{\alpha\beta}\right)\frac{1}{
\,m_g^2\,-\,p^2\,-
i\epsilon}\,,
\label{5D}
\eeq
where
\begin{equation}
\tilde\eta_{\mu\nu}\,\equiv \,\eta_{\mu\nu}\,-\,
\frac{p_\mu p_\nu}{m_g^2}\,.
\label{singm}
\end{equation}
We draw the reader's attention to
 the $1/m_g^4$, $1/m_g^2$ singularities of the above propagator.
The fact of their occurrence  will be important in what follows.

It is the difference in the  numerical coefficients in front of the $\eta_{\mu\nu}
\eta_{\alpha\beta}$ structure in the massless vs. massive propagators 
(1/2 versus 1/3) that  leads to the famous perturbative
discontinuity \cite{Iwa,Veltman,Zakharov}.
No matter how small the graviton mass
is, the predictions are substantially  different
in the two cases. The structure (\ref {5D}) gives
rise to contradictions with observations.

However, as was first pointed out in Ref.~\cite {Arkady},
this discontinuity could be an artifact of relying on the
tree-level perturbation theory which, in fact,
badly breaks down at a  higher nonlinear level  \cite {Arkady,DDGV}.
One should note that the discontinuity does no appear on curved backgrounds
\cite{Ian,Porrati} --- another indication of the spurious nature
of the ``mass discontinuity phenomenon." 

To see the failure of the perturbative expansion
in the Newton constant $G_N$ one could examine the Schwarzschild
solution of the model (\ref {PF}), as  was done in Ref.~\cite{Arkady} (see also
\cite{cimento,Salam,Kogan}). However, probably the easiest way  to understand the 
perturbation theory breakdown  is through examination of 
 the tree-level trilinear graviton vertex diagram. 
At the nonlinear level we have two
extra propagators which could provide a singularity in $m_g$ up to 
$1/m_g^8$.

Two leading terms, $1/m_g^8$ and $1/m_g^6$, do not contribute  \cite{DDGV},
so  that the worst  singularity is $1/m_g^4$.
This is enough to  lead to the perturbation theory breakdown. For a
Schwarzschild source of mass $M$ the breakdown  happens \cite{Arkady,DDGV} at
the scale 
$$
\Lambda_m\sim m_g (Mm_g/\mpl^2)^{-1/5}\,.
$$  
The result can also be understood in terms of
interactions of longitudinal polarizations of the massive graviton which 
become strong \cite{AGS}. For the gravitational sector 
{\em per se}, the corresponding scale $\Lambda_m$  reduces to \cite{AGS} 
$$
m_g  (m_g/\mpl)^{-1/5}\,.
$$
If one uses the freedom associated with possible addition of  higher nonlinear
terms,  one can make \cite {AGS} the breaking scale  as large as 
$$
m_g/ (m_g/\mpl)^{1/3}\,.
$$
(Note that at the classical level the 
strong-coupling problem of the PF gravity can be evaded by summing up 
tree-level nonlinear diagrams \cite{Arkady,DDGV}. 
To determine whether the problem is present at the quantum level, 
one must perform perturbations on a stable background; 
however, the Minkowski-space background is not stable for 
the PF gravity, with the instability setting  
in almost instantaneously \cite {GG}. For recent discussions of massive 
gravity see Refs. \cite {CedricM}, \cite {Grojean}.).

Summarizing, in the diagrammatic language  the reason
for the precocious breakdown of perturbation theory
can be traced back to the infrared terms in the propagator (\ref {5D}) 
which  scale as
\beq
{p_\mu \,p_\nu\over m_g^2} \,.
\label{sing}
\eeq
These terms do not manifest themselves at the linear level;
however,  they do contribute to nonlinear vertices creating problems
in the perturbative treatment of massive gravity already in a classical theory.

We will see momentarily  that   similar problems are totally absent
in the propagator of the model (\ref {1}). For illustrational purposes
it is sufficient to treat the $N=2$ case.  All necessary
calculations were carried out  in  
Sect.~\ref{potg}. Therefore,  here we just assemble  relevant answers. 

For $N=2$  and $b>1/3$ we find
\beq
{p_\mu \,p_\nu\,D(p,y) \over 2m_c^2\,-\,(3b-1)p^2\,D_0(p)\,+\,i\e} \,.
\label{6Dpp0}
\eeq
In the limit $m_c\to 0$ the above expression, as opposed
to Eq.~(\ref {sing}), is {\em regular}. 
Similar calculations can be done in the $N>2$ case. The results is
proportional to
\beq
{\,p_\mu \,p_\nu\,D(p,y) \over (2+N) \,u^N\,-\,k_N\,p^2\,D_0(p)\,+\,i\e} \,,
\label{Dpp0}
\eeq
which is also regular in the $m_c\to 0$ limit
where it  approaches the 4D expression.
Therefore, we conclude  that there is no reason to
expect any breaking of  perturbation theory in the model (\ref {1})
 below  the scale of its UV cutoff.

\vspace{1mm}

If $b<1/3$ and $N=2$  we find,  by the same token,
\beq
{p_\mu \,p_\nu\,D(p,y)\over 2} \left (  {
1 \over 2m_c^2\,-\,(3b-1)p^2\,D_0(p)\,+\,i\e} \,+\,
(\e\to -\e) \right )\,.
\label{6Dpp}
\eeq
Again, in the limit $m_c\to 0$ the above expression, in contradistinction with
 Eq.~(\ref {sing}), is regular. Moreover, in this limit
(and at $y=0$) it approaches the 4D expression,
in a particular gauge.
Analogous calculations can be readily done in the $N>2$ and $b<(2N-2)/3N$  
case. The results is
\beq
{p_\mu \,p_\nu\,D(p,y)\over 2} 
\left[
{1 \over (2+N) u^N - k_N\,p^2\,D_0(p)+ i\e}
+ (\e\to -\e)
\right] .
\label{Dpp}
\eeq
This expression is also regular in the $m_c\to 0$ limit
where it  arrives at the correct 4D limit.
We conclude therefore,  that in the general case  there is no reason to
expect any breaking of  perturbation theory in the model (\ref {11})
below  the scale of its UV cutoff. Note that the 
expressions (\ref {6Dpp}) and (\ref {Dpp}) are singular for 
small Euclidean momenta $p^2\sim - m_c^2$. By construction this singularity has no 
imaginary part and there is no physical state associated with it. One might expect 
that this singularities will be removed after the loop corrections are taken into account 
in a full quantum theory. These considerations are beyond the scope of the present work.
 
An analogy with the Higgs mechanism for non-Abelian gauge fields
is in order here. For massive non-Abelian gauge fields
nonlinear amplitudes violate the unitarity bounds at the scale set by the 
gauge field mass.
This disaster is cured  through the  introduction of  the Higgs field.
Likewise, nonlinear amplitudes
of the 4D massive gravity (\ref {PF}) blow up
at the scale $\Lambda_m$. The unwanted explosion is canceled 
at the  expense of introducing an infinite number of the
Kaluza-Klein fields in  (\ref {1}).

\section{Discussion and conclusions}
\label{dac}

In the present work we studied the model (\ref {11}) 
of the brane-induced gravity in codimensions two and higher. This model 
has   stringent and testable predictions. 
Gravity is modified at short distances, of
the order of $\sim 0.1$ mm or so, and simultaneously, 
at ultra-large distances,  of the order of $\sim 10^{29}$ mm, 
give or take an order of magnitude. 
The short-distance modification  can be tested in table-top 
gravitational experiments \cite{mm,Savas}.
Modification of gravity at a millimeter scale and its relation 
to the cosmological constant problem was first discussed in 
Ref. ~\cite{Sundrum}.  

The modification of gravity at a millimeter scale 
in the present model (\ref {1}) is a consequence  of 
the large-distance modification and {\it vise versa}.
These are two faces of one and the same phenomenon.
However, we should point out that technically and conceptually
the approaches to the cosmological constant problem  
discussed in Ref.~\cite{Sundrum} on the one hand, 
and  Refs.~\cite{DG,DGS} on the other,
are rather different --- the former relies on the 
short-distance (UV) modification of gravity, 
while the latter is entirely based on the large-distance (IR) 
modification of gravity. 

The large distance modification of gravity can manifest itself 
in cosmological solutions. The case $b=0$ and $a=1$ 
seems to be most interesting for these purposes.
As we argued in the present work, it leads to apparent 
violations of 4D causality which  could  manifest 
themselves  at the scales of the order of today's Hubble scale. 
Manifestations of acausality  might be tested   in cosmological observations.
In particular, such an acausal theory might be the 
reason behind  the smallness of the observable space-time curvature
\cite {ADDG}.

It is instructive to point out how the $b=0$, $a=1$ model evades a 
well-known no-go theorem for massive gravity \cite {Veltman}. 
Let us first briefly recall the theorem. A 4D massive graviton has 
extra polarizations one of which couples to sources in the 
leading order in a weak field.  The additional attraction due to this 
polarization is  observationally unacceptable and has to be canceled. 
This can be achieved by introducing a ghost that gives rise to compensating 
repulsive force \cite {Veltman}. Hence, either one ends up with a theory that has 
a ghost or with a theory that has no ghosts but is phenomenologically unacceptable. 
This is the essence of the no-go theorem \cite {Veltman}.  
The theorem can easily be generalized for a theory with an infinite number of 
states \cite {DGPnogo}.  In the latter case, in order to obtain 
a phenomenologically acceptable theory of a massive graviton at observable distances 
one should  give up positivity of the spectral function  in the dispersion relation
for the corresponding Green's function. This would violate unitarity of the model.
However, the above argument assumes 4D analyticity the consequence of which is the 
existence of a spectral representation for the Green's function. In our case 4D 
analyticity is violated, and so are the conditions of the no-go theorem.

We would like to point out a certain common feature 
with $(2+1)$ topologically massive  gauge/gravity theory \cite {Top} 
where the large distance interactions are also power-like.
  
Finally, we would like to emphasize that the models 
(\ref {1}) and (\ref {11}) give  rise to a gravitational analog of the 
Higgs mechanism in the following 
sense --- the effective graviton-mediating interaction  is massive, and,
nevertheless, the growth  of the nonlinear amplitudes is softened 
at the expense of having an infinite number of fields.
This phenomenon can be rather transparently understood from the
standpoint   of the KK modes.
The manifest reparametrization invariance is a convenient book-keeping 
tool in this case for determining whether or not 
the amplitudes blow up. The reparametrization invariance at each KK level is 
maintained at the same KK level {\em only} in the 
linearized approximation. Nonlinear effects mix distinct 
KK levels under the reparametrization transformations \cite{Nappi}, \cite{Stelle}. 
Hence, if the KK tower is truncated 
at some finite level,  
the breakdown of perturbation theory 
in nonlinear  diagrams is inevitable.  However, if the infinite totality
of the KK modes are kept, as in Eq.~(\ref {1}), the softening of the 
amplitudes should be expected. The present work fully confirms this 
expectation. In light of this finding,
it would be interesting to discuss  
the strong coupling issue  in nonlinear interactions
of the 5D DGP model studied in Refs. \cite {DDGV,Luty,Rub}. 
In particular, the immediate task is to understand  
whether this is a problem of peculiar perturbation theory, 
as advocated in Ref. \cite{DDGV}, or the problem inherent 
to the model itself~\cite{Luty,Rub}. Already from the equations of the 
present work it is clear that the $N=1$ DGP model with $b=1$ in (\ref {11})
has no strong coupling problem.  The other possibilities are currently under  
investigation;  the answers will be reported elsewhere.

\vspace{5mm}

{\bf Acknowledgments}

\vspace{3mm}  

We would like to thank Ignatios  Antoniadis, Leonardo Giusti, 
Gia Dvali, Massimo Porrati, 
Arkady Vainshtein, Pierre Vanhove and  Gabriele Veneziano 
for useful discussions.  A significant part of this work was
carried out while both authors were at CERN. 
We are grateful to the CERN Theory Division for kind hospitality.
The work  of M.S. was
supported in part by DOE grant DE-FG02-94ER408.

\end{document}